\documentclass[sigconf,nonacm,screen]{acmart}
\usepackage{multirow}
\usepackage{colortbl}
\usepackage{algorithm}
\usepackage{algorithmic}
\usepackage{listings}
\usepackage{xspace}
\newcommand{\ourproblem}{\texttt{IntentGEO}\xspace}
\lstdefinestyle{promptstyle}{basicstyle=\ttfamily\scriptsize,breaklines=true,
  columns=fullflexible,keepspaces=true,
  literate={—}{{---}}1 {’}{{'}}1 {“}{{``}}1 {”}{{''}}1 {✓}{{[OK]}}1 {✗}{{[X]}}1}

\setcopyright{cc}
\setcctype[4.0]{by}
\AtBeginDocument{}

\begin{document}

\title{\ourproblem: Intent-aware Generative Engine Optimization for Recommendation}

\author{Runlin Lei}
\affiliation{\institution{Renmin University of China}\city{Beijing}\country{China}}
\email{runlin\_lei@ruc.edu.cn}
\author{Shiqi Zhang}
\authornote{Zhewei Wei and Shiqi Zhang are the corresponding authors.}
\affiliation{\institution{National University of Singapore}\country{Singapore}}
\email{shiqi@pyrowis.ai}
\author{Xiaokui Xiao}
\affiliation{\institution{National University of Singapore}\country{Singapore}}
\email{xkxiao@nus.edu.sg}
% TODO: add this author's email if desired for the public preprint.
\author{Zhewei Wei}
\authornotemark[1]
\affiliation{\institution{Renmin University of China}\city{Beijing}\country{China}}
\email{zhewei@ruc.edu.cn}
\renewcommand{\shortauthors}{Lei et al.}

\begin{abstract}
Large language models (LLMs) increasingly shape recommendations across chatbots, e-commerce, and user-generated content (UGC) platforms, creating demand for Generative Engine Optimization (GEO) to improve item visibility in generated responses.
Existing GEO studies are mainly \textit{content-centric}, optimizing visibility under general-purpose queries.
However, a shared item description usually serves users with different needs and decision goals, resulting in its visibility gains being dependent on the request.
To investigate how changing intent affects visibility, we formulate \ourproblem to evaluate shared rewrites across need specificity and decision progression.
We revisit existing GEO methods and find that their visibility gains vary across user intents.
More critically, learning-based prompt optimization can underperform its handwritten seed on average.
Tracing these updates reveals \textit{prompt drift} from explicit recommendation instructions toward generic promotional tactics.
To address this drift, we propose \textbf{AMO} (\emph{Anchored Meta-Optimization}), which preserves recommendation instructions in a fixed core while learning appended strategies from citation feedback.
Across four domains and ten intent states, AMO improves average citation-rank gain by 35\% over the strongest prior learning-based baseline and leads on nine states. 
Ablations show that learned additions improve on the fixed core alone and that AMO outperforms full-prompt optimization initialized from that core.
It also achieves the highest gains in an agentic recommendation pipeline and, after separate adaptation, on Doubao, a commercially deployed, closed-source online assistant. 
These findings highlight user intent as a key dimension of GEO evaluation and support anchored adaptation across recommendation needs and decision stages.

\end{abstract}



\begin{teaserfigure}
\includegraphics[width=\textwidth]{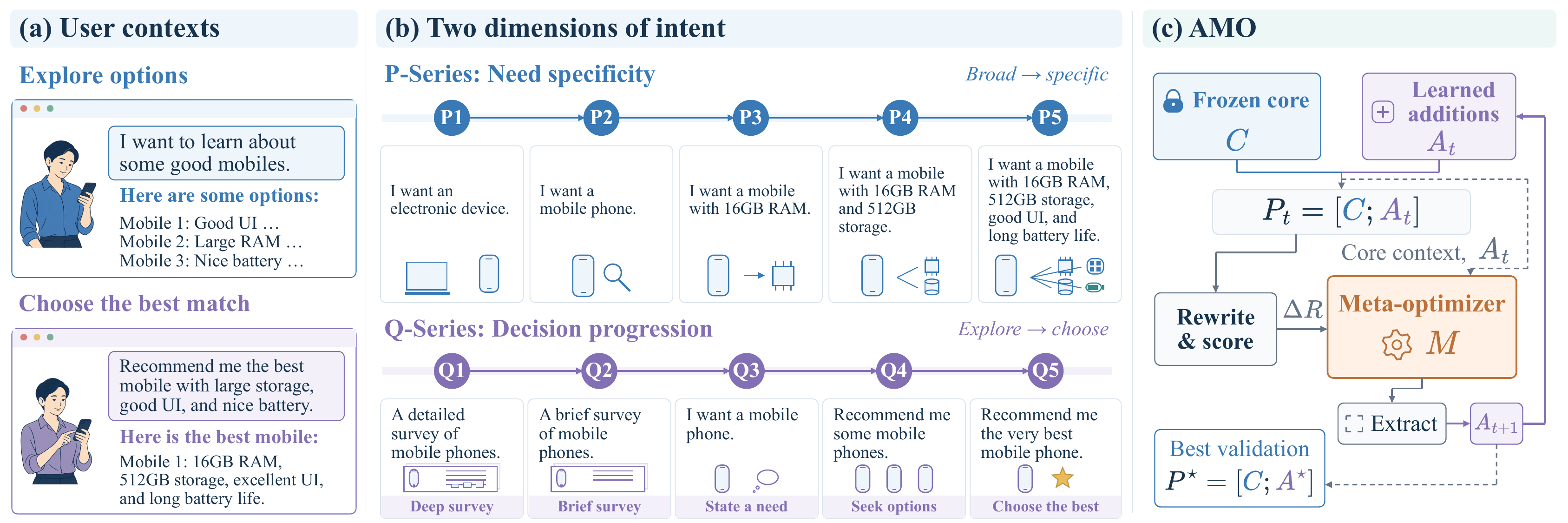}
\caption{\textbf{Intent-aware recommendation GEO.} (a) User needs vary in specificity and decision stage. (b) P/Q queries probe these two dimensions. (c) AMO preserves a recommendation core while learning appended strategies.}
\Description{Three panels show recommendation contexts, five levels of need specificity and decision progression, and an anchored meta-optimization loop that learns additions while keeping core recommendation instructions fixed.}
\label{fig:intent_aware_queries}
\end{teaserfigure}

\maketitle

\section{Introduction}
Recent advances in large language models (LLMs) and AI chatbots such as ChatGPT are reshaping how users seek information on the Web, shifting search behavior from keyword-based retrieval to intent-driven questions answered by LLMs. 
For example, in 2024 Gartner projected that traditional search engine volume would decrease by 25\% by 2026 as users migrate to AI chatbots~\citep{gartner_predicts_2024}. 
This shift is also changing search and recommendation interfaces across search engines, e-commerce platforms, and user-generated content (UGC) platforms, from ranked lists to direct answers generated by integrated generative engines~\citep{google_ai_mode, amazon_alexa_shopping, reuters_qwen_taobao_2026}. 
As generated responses become a new source of organic traffic, brand owners increasingly need ways to improve their visibility in these answers.

Generative Engine Optimization (GEO) studies how to rewrite item content so that generative engines are more likely to use it when answering relevant user queries~\cite{geo}.
Commonly, such rewriting is evaluated within a fixed candidate document pool by measuring changes in source visibility, modeling how an LLM selects and presents retrieved content in its generated answer~\cite{geo,c-seo,ragdoll}.
Existing methods pursue this objective through content-centric improvements in fluency and authority~\cite{c-seo}, black-box optimization~\cite{black-hat-geo}, and learned rewriting strategies~\cite{meta-optimization,autogeo}.

In recommendation, however, users seeking the same type of item may ask very different questions, depending on their product requirements and stage in the decision process~\cite{hirsch2020query,chen2021reformulation,lemon2016understanding}.
For example, a user's need may range from a broad category-level request, such as \textit{``best mobile phone''}, to a specific constraint-driven inquiry, such as \textit{``mobile phone with more than 16GB RAM''}.
At the same time, the user's decision stage can range from an exploratory market survey, such as \textit{``survey the mobile phone market''}, to seeking a direct recommendation, such as \textit{``recommend the best mobile phone''}.
These variations change which item attributes matter and how an assistant uses them to justify a recommendation.
As providers have to publish a shared description before knowing each user's request, the gains from GEO under one query may not persist as needs become more specific or the requested choice narrows.

To study GEO under these variations, we formulate \ourproblem along two dimensions: \textit{need specificity} and \textit{decision progression}.
Specifically, we characterize user intent through \textit{need specificity} and \textit{decision progression}.
Our need-specificity queries narrow from broad categories to attributes documented for the target.
Decision progression captures the shift from exploring options to seeking a single best recommendation.
To examine whether rewriting gains persist as these requests change, we compare each shared rewrite with the original description while keeping the target item and candidate pool fixed.
As a reference for recommendation, we design RecGEO, a simple handwritten rewriting prompt.
Across both request families, existing GEO methods yield query-dependent gains and can lose visibility as needs become more specific, while RecGEO already achieves competitive aggregate performance.
Moreover, learning-based baselines initialized with RecGEO remain below this handwritten reference, motivating a closer examination of how optimization changes recommendation-specific instructions.
  
To understand this decline, we examine the optimization process and find that learned prompts can drift away from the seed's recommendation focus.
This \textit{prompt drift} motivates \textbf{AMO} (Anchored Meta-Optimization), which separates fixed recommendation requirements from learned rewriting strategies. 
During learning, citation feedback guides these refinements while the core instructions remain intact. 
Experiments show that AMO improves visibility across both intent dimensions, and learned refinements improve aggregate visibility over the fixed core alone. A full-prompt variant initialized with the same core also yields lower gains.
We further reuse these rewrites in an agentic recommendation workflow and separately adapt AMO using citation feedback from the commercially deployed Doubao assistant. 
The results show that AMO achieves the largest mean citation-rank gains in both settings.

Our contributions can be summarized in four aspects.
\begin{enumerate}
\item We formulate \ourproblem to evaluate shared rewrites under changing recommendation needs and decision goals.

\item We find intent-dependent GEO gains and show that prompt optimization can underperform its seed on average.

\item We propose \textbf{AMO}, which learns additions to a fixed recommendation core and improves average visibility across the evaluated wordings and backbones.

\item We extend GEO evaluation to agentic recommendation and on Doubao, a commercially deployed online assistant.
\end{enumerate}

\section{Problem Setup}
\label{sec:preliminary}

Generative engines retrieve documents relevant to a query and use them as context to generate a cited answer~\cite{geo}. 
GEO benchmarks evaluate content rewriting at this generation stage, keeping the candidate pool fixed while measuring changes in source visibility~\cite{c-seo,ragdoll}.
In this paper, we follow this setting and let $G$ denote a generative engine that takes a query $q$ and a candidate document set $\mathcal{D}_q=\{d_1,\ldots,d_n\}$ as input and returns a cited answer $y=G(q,\mathcal{D}_q)$.
Let $\pi(y)=(c_1,\ldots,c_m)$ denote the ordered list of unique citations in $y$, indexed by first appearance, where $c_j\in\{1,\ldots,n\}$. The citation rank of a document $d_i$ is:
\[
\operatorname{rank}_y(d_i)=
\begin{cases}
j, & c_j=i,\\
n+1, & d_i \notin \pi(y).
\end{cases}
\]
To optimize the target item's visibility, GEO rewrites the content of $d_t$ into $\tilde d_t$. 
Note that only the target document is changed, while candidate identities, input order, and all other document contents remain fixed across methods.
We adopt the \textbf{white-hat} setting in C-SEO Bench~\cite{c-seo}, where the optimizer assumes no access to model parameters, and the optimized document must remain natural, factual, and usable in realistic recommendation scenarios.
For each instance, we randomly select an eligible target $d_t$ and keep it fixed across methods and intent states (Appendix~\ref{sec:appendix-data}). 
For each evaluated request, $y^{(0)}$ and $y^{(1)}$ denote answers with the original and rewritten target descriptions, respectively.
We then define
\[
r^{(0)}=\operatorname{rank}_{y^{(0)}}(d_t), \qquad
r^{(1)}=\operatorname{rank}_{y^{(1)}}(d_t).
\]
Following C-SEO Bench, we measure the improvement in the target item's visibility using Delta Rank:
\[
\Delta R=r^{(0)}-r^{(1)}.
\]
Because an uncited target receives rank $n+1$, $\Delta R$ combines citation inclusion with changes in first-citation order.
A positive value means the rewrite makes the target appear earlier in the citation list or brings an uncited target into the answer.
Appendix~\ref{app:robustness} reports additional visibility metrics, including Pass@$K$ and MRR@$K$.

\section{The \ourproblem Problem}
\label{sec:motivation}

\subsection{Intent-aware Queries in Recommendation}
Recommendation queries vary with user intent~\cite{broder2002taxonomy, jannach2024survey_intent_aware_recommendation}.
Such variation occurs across users and within the same person's search for a product. Prior research shows that users often begin with uncertainty and a vague query, and progressively move toward a clearer and more focused formulation as needs are specified~\cite{kuhlthau1991inside}. 
At the same time, research shows that requests with similar surface forms may serve different underlying goals~\cite{rose2004understanding_user_goal}. In practice, these goals vary with the user’s stage in the decision journey, such as prepurchase exploration versus purchase-time selection~\citep{lemon2016understanding}.

As user intent is central to recommendation evaluation~\citep{herlocker2004evaluating_user_task}, we examine this heterogeneity in GEO through \textbf{intent-aware queries}, varying the requirements and requested choice within the same recommendation instances.

\subsection{Design of the \ourproblem Problem}

To make these aspects testable, we formulate \ourproblem along two dimensions of a request: \textit{need specificity}, the detail of its requirements, and \textit{decision progression}, the requested stage from exploration to a final choice.
These dimensions capture two central aspects of recommendation: identifying items that meet a user's needs and helping users choose among them.
The P-series and Q-series instantiate these dimensions while sharing the original query's topic, candidate pool, and target document.
Figure~\ref{fig:intent_aware_queries} illustrates the progression.

\paragraph{P-Series: Need Specificity.}
The P-series varies requirements across five levels. \textbf{P1} is a broad domain- or category-level keyword; \textbf{P2} adds a topic in a noun-phrase request, such as ``a book about Arctic exploration saga''.
To generate \textbf{P3--P5}, the generator receives the original topic, target item, and candidate descriptions, then expresses progressively more source-supported target attributes as user requirements.
This target-conditioned progression makes the item's fit increasingly explicit, testing whether rewriting still adds visibility as the match strengthens.
For example, users may express such feature-based requirements before choosing a particular item, as when a shopper seeks a lightweight laptop with long battery life without naming a model.

To assess rewriting under these requirements, we compare the original and rewritten descriptions under the same request at each level.
A negative gain therefore indicates that rewriting has reduced the target's visibility despite the evidence of its fit in the source.

\paragraph{Q-Series: Decision Progression.}
The Q-series uses templates to vary the recommendation task requested of the assistant. 
\textbf{Q1} requests a comprehensive survey, \textbf{Q2} a brief survey, \textbf{Q3} the topic-level anchor, \textbf{Q4} multiple recommendations, and \textbf{Q5} the single strongest recommendation. P2 and Q3 share wording but receive independent answers.
These requests narrow selection from exploring options to making a final choice.

Under each request, we compare the shared rewrite with the unmodified target using the same candidate pool. 
State-wise gains show where rewriting helps, while family averages summarize its performance across requests.
Together, the families define ten evaluation slots, with the shared anchor counted once in each family average. Appendix~\ref{app:prompt} describes query construction and provides the templates.

\begin{figure*}[t]
\centering
\includegraphics[width=0.88\textwidth]{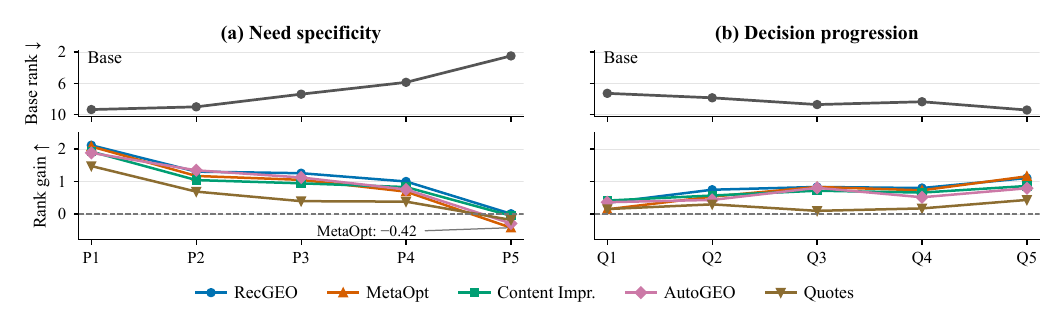}
\caption{Base visibility and GEO effectiveness vary across recommendation requests. Top: Base citation rank (lower is better). Bottom: rank gains of the five strongest original baselines (higher is better); negative gains indicate lost visibility.}
\Description{Two columns compare need specificity P1 through P5 and decision progression Q1 through Q5. Base rank decreases across P levels, while the five GEO baselines lose gain and most become negative at P5. Q-series gains remain positive but vary by intent state.}
\label{fig:revisit-rank-curves}
\end{figure*}

\subsection{Experimental Setup}
\label{sec:experiment-setup}
To examine whether visibility gains persist as user intent changes, we compare prior GEO methods across the ten P/Q states.

\paragraph{Datasets.}
We evaluate on four recommendation datasets: Books, Retail, and Videogames from C-SEO Bench~\citep{c-seo}, and Ragdoll~\cite{ragdoll}.
The original queries follow C-SEO Bench's topic-level template (e.g., ``a book about Arctic exploration saga'').
We sample 50 held-out instances per domain and expand each into the ten P/Q states, yielding 200 test instances and 2,000 evaluation requests.
This design compares intent states within the same recommendation instances.

For prompt optimization, we use a separate set of 574 training and 75 validation examples from the three C-SEO domains.
Ragdoll contributes only test instances, allowing us to evaluate transfer to a collection unseen during prompt training and selection.
Validation uses topic-level requests.
Training uses either the shared P2/Q3 topic-level request or, in progressive training, a sampled P/Q state for each example.
We provide the full summary of data statistics in Table~\ref{tab:data-splits} in Appendix~\ref{sec:appendix-data}.

\paragraph{Baselines.}
We compare heuristic and learning-based GEO methods.
The heuristic baselines comprise the ten rewriting methods from C-SEO Bench~\cite{c-seo}, listed in Table~\ref{tab:main-results-detailed}, and RecGEO, a handwritten recommendation-oriented prompt.
The C-SEO methods range from targeted edits, such as Fluency and Citations, to holistic Content Improvement (Content Impr.).
RecGEO instructs the rewriter to highlight distinctive strengths and recommendation-worthy evidence while preserving the target item's identity and factual content.
We ensure its prompt length to be comparable to the other training-free baselines.

Learning-based methods derive rewriting strategies from engine behavior.
AutoGEO~\citep{autogeo} applies learned engine-preference rules to rewriting.
We use its released e-commerce rule set.
MetaOpt~\citep{meta-optimization} iteratively updates the rewriting prompt using rank feedback, starting from Content Impr.
To test whether initialization or intent coverage improves adaptation, we vary MetaOpt's seed prompt and training queries.
The ``-Rec'' variants replace the seed with RecGEO, while ``-Prog'' replaces topic-level training queries with sampled P/Q states.
These choices yield MetaOpt, MetaOpt-Rec, MetaOpt-Prog, and MetaOpt-Rec-Prog under shared optimization settings.

\paragraph{Evaluation.}
We use \texttt{gpt-4o-mini}~\cite{gpt-4o-mini} as the backbone unless otherwise specified.
AMO and MetaOpt share training and validation splits.
Training updates a rewriting prompt, and validation selects a single candidate without test answers.
The selected prompt generates three independent rewrites of each test target.
We evaluate each saved rewrite once under each of the ten P/Q requests, keeping the description fixed as the assistant's task changes.
We report means in the main text and standard deviations across rewrite variants in Appendix~\ref{app:full-results}.
Appendix~\ref{app:param} details the configuration.

\subsection{Results under \ourproblem}

We revisit prior GEO methods under the proposed \ourproblem problem.
The results of the top-five GEO baselines are shown in Figure~\ref{fig:revisit-rank-curves}, with the full results deferred to Table~\ref{tab:main-results-detailed}.

\paragraph{Base performance trends vary across intent families.}
Base visibility responds to the evidence and choice specified in the request.
Along the P-series, the target's mean rank improves from 9.33 at P1 to 2.50 at P5 as requests add attributes that the target supports.
The progression gives the assistant increasingly concrete reasons to consider the same target among the candidates.
In the Q-series, mean Base rank changes from 7.27 under the Q1 survey request to 9.39 under Q5's single-best recommendation request.
As the requested choice narrows, the target needs more distinctive recommendation evidence to stand out.
The two dimensions thus place different demands on the same description.

\paragraph{Many GEO methods lose visibility under closely matched requests.}
Closely matched requests expose a practical risk of GEO.
Most prior methods yield negative gains at P5, where the requested attributes are documented for the target.
For these methods, rewriting lowers visibility on average even when the source already gives the assistant concrete reasons to recommend the item.
Along the P-series, declining gains accompany more specific requests and stronger Base rankings, which leave less room for improvement.
These patterns motivate evaluating a shared rewrite under both broad and closely matched recommendation requests.

Across the Q-series, the top-five methods retain positive gains, with their effectiveness varying by the requested recommendation choice.
The handwritten RecGEO prompt is already competitive with the learning-based baselines, making it a promising seed for further optimization.
We therefore initialize MetaOpt with RecGEO and test whether iterative updates guided by citation feedback can improve its gains.

\begin{table}[t]
\centering
\caption{
RecGEO and four MetaOpt variants.
The ``-Rec'' variants use RecGEO as the seed prompt, and ``-Prog'' variants replace each original training query with a randomly sampled P- or Q-series variant.
}
\label{tab:meta-variant-comparison}
\begin{small}
\begin{tabular}{lcccc}
\toprule
Method & P5 & P-Avg & Q-Avg & All \\
\midrule
RecGEO seed & \textbf{+0.01} & \textbf{+1.14} & \textbf{+0.77} & \textbf{+0.95} \\
\midrule
MetaOpt  & -0.42 & +0.91 & +0.68 & +0.80 \\
MetaOpt-Rec & -0.66 & +0.79 & +0.56 & +0.67 \\
MetaOpt-Prog & -0.42 & +0.98 & +0.68 & +0.83 \\
MetaOpt-Rec-Prog & -0.58 & +0.85 & +0.57 & +0.71 \\
\bottomrule
\end{tabular}
\end{small}
\end{table}

\paragraph{MetaOpt can weaken an effective recommendation seed.}

To test this expectation, we vary MetaOpt's seed and training queries (Table~\ref{tab:meta-variant-comparison}).
Yet MetaOpt-Rec lowers the seed's overall gain from $0.95$ to $0.67$.
Both RecGEO-seeded variants also fall below their Content Impr.-seeded counterparts.
P/Q-mixed training offers only small improvements, leaving all variants below the handwritten seed.
We next examine how MetaOpt changes an already effective recommendation prompt to understand this unexpected degradation.

\section{Method}
Motivated by MetaOpt-Rec's performance drop, we ask how optimization changes an effective recommendation seed (\textbf{RQ1}) and how to retain recommendation instructions during prompt adaptation (\textbf{RQ2}).
We develop AMO from this analysis and evaluate whether it improves visibility across user intents (\textbf{RQ3}).
We then examine which components contribute to its gains, how well they transfer, and whether the rewrites retain acceptable quality (\textbf{RQ4}).

\subsection{Prompt Drift in Meta-Optimization}
\label{sec:core-repair}

\paragraph{Diagnostic design.}
MetaOpt can revise the recommendation instructions and rewriting strategies in the same prompt.
To examine how it uses this flexibility (\textbf{RQ1}), we track four literal instruction markers.
Complete factual retention and a recommender's voice represent the seed's explicit priorities, while keywords and calls to action capture recurring promotional directions in the updates.
Table~\ref{tab:prompt-drift} records their presence in the seed and validation-selected prompt and across 36 saved updates from 3 epochs with 12 batches each.
Appendix~\ref{app:drift-markers} gives the matching rules.

\begin{table}[htb]
\centering
\caption{Literal instruction markers before and after MetaOpt-Rec optimization. ``Updates'' counts saved prompts containing each marker.}
\label{tab:prompt-drift}
\begin{tabular}{@{}lccc@{}}
\toprule
Instruction marker & Seed & Selected & Updates \\
\midrule
Full retention & Yes & No & 3/36 \\
Recommender voice & Yes & No & 0/36 \\
Keywords & No & Yes & 36/36 \\
Call to action & No & Yes & 36/36 \\
\bottomrule
\end{tabular}
\end{table}

\begin{table*}[!t]
\centering
\caption{
Main benchmark results in \ourproblem. The Base row reports citation ranks; other rows report rank gains ($\uparrow$).
Best gains are \textbf{bold} with a light red background; second-best gains are \underline{underlined} with a gray background. Ties follow displayed values.
}
\label{tab:main-results-detailed}
\begin{tabular}{lrrrrrrrrrrrrr}
\toprule
\multirow{2}{*}{Method} & \multicolumn{5}{c}{Need specificity} & \multicolumn{5}{c}{Decision progression} & \multicolumn{3}{c}{Average} \\
\cmidrule(lr){2-6}\cmidrule(lr){7-11}\cmidrule(l){12-14}
 & P1 & P2 & P3 & P4 & P5 & Q1 & Q2 & Q3 & Q4 & Q5 & P & Q & All \\
\midrule
Base & 9.33 & 8.98 & 7.37 & 5.87 & 2.50 & 7.27 & 7.83 & 8.70 & 8.33 & 9.39 & 6.81 & 8.30 & 7.56 \\
\midrule
Authoritative & +1.36& +0.54& +0.21& +0.40& -0.12& +0.00& +0.17& +0.14& +0.15& +0.35& +0.48& +0.16& +0.32 \\
Statistics & +1.19& +0.48& +0.21& +0.24& -0.18& -0.04& +0.19& -0.14& +0.07& +0.36& +0.39& +0.09& +0.24 \\
Citations & +1.20& +0.56& +0.28& +0.36& -0.14& +0.05& +0.19& +0.11& +0.21& +0.30& +0.45& +0.17& +0.31 \\
Fluency & +1.19& +0.71& +0.34& +0.45& -0.27& +0.14& +0.25& +0.28& +0.24& +0.27& +0.48& +0.23& +0.36 \\
Unique Words & +0.99& +0.35& +0.03& +0.01& -0.26& -0.15& +0.12& -0.05& +0.07& +0.06& +0.22& +0.01& +0.12 \\
Tech. Terms & +1.19& +0.52& +0.24& +0.23& -0.22& +0.03& +0.21& +0.09& +0.07& +0.21& +0.39& +0.12& +0.26 \\
Simple Lang. & +1.18& +0.53& +0.17& +0.09& -0.37& +0.07& +0.22& +0.07& +0.15& +0.13& +0.32& +0.13& +0.22 \\
Quotes & +1.48& +0.69& +0.40& +0.38& -0.18& +0.15& +0.30& +0.10& +0.17& +0.43& +0.55& +0.23& +0.39 \\
LLM Guid. & +1.09& +0.39& +0.28& +0.23& -0.70& -0.52& -0.28& +0.01& -0.16& +0.29& +0.26& -0.13& +0.06 \\
Content Impr. & +1.92& +1.04& +0.94& +0.83& -0.05& \cellcolor[gray]{0.94}\underline{+0.42}& +0.57& +0.72& +0.66& +0.86& +0.94& +0.65& +0.79 \\
RecGEO & \cellcolor[gray]{0.94}\underline{+2.12}& +1.31& \cellcolor[rgb]{1.00,0.91,0.90}\textbf{+1.26}& \cellcolor[gray]{0.94}\underline{+1.00}& \cellcolor[gray]{0.94}\underline{+0.01}& +0.35& \cellcolor[gray]{0.94}\underline{+0.75}& +0.83& \cellcolor[gray]{0.94}\underline{+0.80}& +1.11& \cellcolor[gray]{0.94}\underline{+1.14}& \cellcolor[gray]{0.94}\underline{+0.77}& \cellcolor[gray]{0.94}\underline{+0.95} \\
\midrule
AutoGEO & +1.88& +1.34& \cellcolor[gray]{0.94}\underline{+1.12}& +0.76& -0.29& +0.36& +0.43& +0.81& +0.52& +0.79& +0.96& +0.58& +0.77 \\
MetaOpt & +2.08& +1.17& +1.05& +0.69& -0.42& +0.14& +0.54& +0.83& +0.73& \cellcolor[gray]{0.94}\underline{+1.16}& +0.91& +0.68& +0.80 \\
MetaOpt-Prog & \cellcolor[gray]{0.94}\underline{+2.12}& \cellcolor[gray]{0.94}\underline{+1.38}& +1.05& +0.79& -0.42& +0.39& +0.49& \cellcolor[gray]{0.94}\underline{+0.85}& +0.68& +0.98& +0.98& +0.68& +0.83 \\
\midrule
\textbf{AMO} & \cellcolor[rgb]{1.00,0.91,0.90}\textbf{+2.44}& \cellcolor[rgb]{1.00,0.91,0.90}\textbf{+1.46}& +1.03& \cellcolor[rgb]{1.00,0.91,0.90}\textbf{+1.13}& \cellcolor[rgb]{1.00,0.91,0.90}\textbf{+0.06}& \cellcolor[rgb]{1.00,0.91,0.90}\textbf{+0.49}& \cellcolor[rgb]{1.00,0.91,0.90}\textbf{+0.78}& \cellcolor[rgb]{1.00,0.91,0.90}\textbf{+1.20}& \cellcolor[rgb]{1.00,0.91,0.90}\textbf{+1.06}& \cellcolor[rgb]{1.00,0.91,0.90}\textbf{+1.59}& \cellcolor[rgb]{1.00,0.91,0.90}\textbf{+1.22}& \cellcolor[rgb]{1.00,0.91,0.90}\textbf{+1.02}& \cellcolor[rgb]{1.00,0.91,0.90}\textbf{+1.12} \\
\bottomrule
\end{tabular}
\end{table*}

\paragraph{Changes in explicit instructions.}
The seed's two instruction markers occur in few saved updates, while both promotional markers appear throughout (Table~\ref{tab:prompt-drift}).
The selected prompt illustrates how the instructions are edited.
``Preserve all factual information'' becomes ``Maintain factual accuracy,'' leaving complete retention less explicit.
It also asks for ``keyword research,'' an ``urgent call to action,'' and ``relevant data, statistics, or testimonials.''
Together, these edits shift the prompt's emphasis from preserving recommendation evidence toward generic promotional appeal.
We refer to this shift in explicit instructions as \textit{prompt drift}.

\paragraph{Motivation for anchoring.}
The \textit{prompt drift} observations motivate keeping recommendation instructions as a stable part of prompt search.
An item's documented attributes provide reasons for a match across different requests, and their presentation can be refined while the instructions to preserve them remain explicit.
AMO therefore fixes a recommendation core and learns additional rewriting strategies through citation feedback.

\subsection{Anchored Meta-Optimization}
\label{sec:amo}
\paragraph{Constrained GEO objective.}
To answer \textbf{RQ2}, we formulate GEO as maximizing citation reward subject to retaining recommendation instructions in the rewriting prompt. 
Let $P$ be the prompt, $\widehat R(P;\mathcal D)$ its empirical citation reward on examples $\mathcal D$, and $\mathcal K$ the required instructions to retain source facts and foreground item-specific recommendation evidence. We seek
\begin{equation}
\begin{aligned}
P^\star &\in \arg\max_P\;\widehat R(P;\mathcal D_{\mathrm{tr}})\\
&\text{subject to }\quad \mathcal K\subseteq\operatorname{Instr}(P),
\end{aligned}
\label{eq:anchored-optimization}
\end{equation}
Here $\operatorname{Instr}(P)$ denotes the directives explicitly supplied in the rewriting prompt.
This \emph{structural regularization} keeps recommendation instructions fixed while citation reward guides the search over rewriting strategies.

\paragraph{AMO implementation.}
\textbf{AMO} learns additions $A$ appended to a fixed recommendation core $C$ (Figure~\ref{fig:intent_aware_queries}).
We obtain $C$ by removing RecGEO's length limit and consolidating its recommendation instructions.
Because $C$ carries $\mathcal K$, each candidate $P=[C;A]$ satisfies the constraint by construction.
Optimization searches for additions with higher citation reward.

We initialize $A_0$ as empty, so search starts from the recommendation core.
At each round, we rewrite a training batch with $P_t=[C;A_t]$ and compute citation-conditioned feedback $\mathcal F_t$ from the engine's answers.
The other candidate descriptions remain unchanged during this comparison.
The meta-optimizer uses this feedback, the update history $H_t$, and a read-only summary of $C$ to propose new additions for the next round of prompt search.
We extract these additions and append them to the unchanged core before evaluating the next candidate.
After each update, we validate $[C;A_{t+1}]$ on held-out topic-level requests and retain the highest-scoring updated prompt.
Algorithm~\ref{alg:amo} summarizes this process; Appendix~\ref{app:amo-optimization} gives the reward and prompt-update details.

\begin{algorithm}[!ht]
\caption{\textbf{AMO}: Anchored Meta-Optimization}
\label{alg:amo}
\renewcommand{\algorithmicrequire}{\textbf{Input:}}
\renewcommand{\algorithmicensure}{\textbf{Output:}}
\begin{algorithmic}[1]
\REQUIRE Training set $\mathcal D_{\mathrm{tr}}$, validation set $\mathcal D_{\mathrm{val}}$, core $C$, optimizer $M$, rounds $T$
\ENSURE Anchored prompt $P^\star=[C;A^\star]$
\STATE $A_0\gets\emptyset$, $A^\star\gets A_0$, $s^\star\gets-\infty$, $H_{-1}\gets\emptyset$
\FOR{$t=0,\ldots,T-1$}
    \STATE $\mathcal B_t\gets\mathrm{NextBatch}(\mathcal D_{\mathrm{tr}})$
    \STATE $P_t\gets[C;A_t]$, $\mathcal F_t\gets\mathrm{Feedback}(P_t;\mathcal B_t)$
    \STATE $H_t\gets\mathrm{Record}(H_{t-1},A_t,\mathcal F_t)$
    \STATE $\widetilde A_{t+1}\gets M(A_t,\mathcal F_t,H_t,\mathrm{summary}(C))$
    \STATE $A_{t+1}\gets\mathrm{ExtractAdditions}(\widetilde A_{t+1})$
    \STATE $P_{t+1}\gets[C;A_{t+1}]$
    \STATE $s_{t+1}\gets\widehat R(P_{t+1};\mathcal D_{\mathrm{val}})$
    \IF{$s_{t+1}>s^\star$}
        \STATE $A^\star\gets A_{t+1}$, $s^\star\gets s_{t+1}$
    \ENDIF
\ENDFOR
\STATE \textbf{return} $[C;A^\star]$
\end{algorithmic}
\end{algorithm}

At test time, the rewriter takes only the target description. AMO reuses its output across requests.

\subsection{Main Results}
After selecting a prompt on topic-level requests, we evaluate its shared rewrites as users add requirements or ask for a more decisive recommendation (\textbf{RQ3}).
Table~\ref{tab:main-results-detailed} reports the resulting rank gains across all ten intent states.
We then examine whether these gains are accompanied by greater citation inclusion and transfer to the unseen Ragdoll collection.
Complete results with standard deviations and Pass@5 are reported in Appendix~\ref{app:full-results}.

\paragraph{AMO improves visibility across diverse intents.}
AMO's overall gain of $1.12$ improves on RecGEO, the strongest heuristic baseline, by $18\%$, and on MetaOpt-Prog, the strongest prior learning-based method, by $35\%$. Paired confidence intervals from instance-level bootstrapping support both aggregate gains (Appendix~\ref{app:uncertainty}). AMO achieves the highest mean gain on nine of the ten states and both intent-family averages (Table~\ref{tab:main-results-detailed}).

\paragraph{Citation inclusion and domain transfer.}
The selected prompt raises the target's top-five citation rate by $13.98$ percentage points over Base (Appendix~\ref{app:robustness}).
In each domain, AMO's mean rank gain exceeds RecGEO and MetaOpt-Prog.
On Ragdoll, which is excluded from training and prompt selection, AMO gains $1.19$ positions, extending the learned strategy to a new collection (Appendix~\ref{app:full-results}).

\subsection{Ablation and Robustness Analysis}
\label{sec:ablation-robustness}

\paragraph{Ablation setup.}
To understand these gains (\textbf{RQ4}), we first examine the roles of the fixed core and learned additions.
\emph{Core only} uses $C$ without learned additions.
AMO edits only $A$, using citation-conditioned reward for feedback and validation after each update.
\emph{Full-prompt optimization} lets MetaOpt edit the entire prompt from the same core, using all-instance $\Delta R$ and validation before each update, including the initial core.
These comparisons ask whether AMO improves on the core and how its complete procedure compares with full-prompt search.
All three share the held-out evaluation protocol and all-instance $\Delta R$ (Appendix~\ref{app:ablation}).

\begin{table}[htb]
\centering
\caption{Rank gains for AMO and its ablation variants.}
\label{tab:core-ablation}
\begin{tabular}{lrrr}
\toprule
\multirow{2}{*}{Method} & \multicolumn{3}{c}{Rank gain $\uparrow$} \\
\cmidrule(lr){2-4}
 & P-Avg & Q-Avg & All \\
\midrule
Core only & 1.15 & 0.80 & 0.98 \\
Full-prompt opt. & 0.84 & 0.64 & 0.74 \\
\midrule
\textbf{AMO} & \textbf{1.22} & \textbf{1.02} & \textbf{1.12} \\
\bottomrule
\end{tabular}
\end{table}

\paragraph{Ablation results.}
\textbf{AMO achieves the highest P and Q averages among the three procedures.}
Learned additions raise the overall gain over Core only from $0.98$ to $1.12$ (Table~\ref{tab:core-ablation}).
The larger improvement occurs on Q queries, with the greatest additional gain at Q5, where the assistant must choose one item.
The Doubao example in Appendix~\ref{app:example} shows an AMO rewrite linking item attributes to reasons for choosing the book.
Full-prompt optimization yields lower gains than both Core only and AMO under its feedback and selection procedure.
Together, these results favor AMO as a complete design for adapting this recommendation seed.

\paragraph{Robustness setup.}
Users can express the same requirements in different ways, and applications may use different answer models.
We therefore test robustness by paraphrasing the P3--P5 requests, rephrasing the core while retaining the learned additions, and transferring the selected prompts to Gemini 2.5 Flash-Lite~\cite{gemini25flashlite} and Qwen3-30B-A3B~\cite{qwen3technicalreport}. 
Each comparison preserves its original targets and candidate pools. Appendix~\ref{app:robustness} gives the settings.

\begin{table}[t]

\centering
\caption{Visibility gains under paraphrased P3--P5 requests.
$\Delta$Pass@5 reports percentage-point gains over Base in the top-five target citation rate.}
\label{tab:query-paraphrase}
\begin{tabular}{lrrr}
\toprule
Method & P3 & P4 & P5 \\
\midrule
\textbf{AMO} & \textbf{14.5} & 8.5 & \textbf{5.5} \\
RecGEO & 11.0 & \textbf{9.0} & \textbf{5.5} \\
MetaOpt-Prog & 10.5 & \textbf{9.0} & 1.5 \\
AutoGEO & 6.0 & 3.5 & 0.5 \\
\bottomrule
\end{tabular}

\end{table}

\paragraph{Robustness results.}
\textbf{AMO remains effective across changes in wording and backbone.} With conversational requests, it leads at P3 and ties RecGEO at P5 (Table~\ref{tab:query-paraphrase}), while gains still decrease as needs become more specific. Rephrasing the core also retains the improvement on both intent families (Table~\ref{tab:anchor-paraphrase}). Finally, AMO achieves the highest overall gains on both transferred backbones (Table~\ref{tab:backbone-sensitivity}). The selected instructions are reused on the transferred backbones without additional prompt optimization.
These checks therefore examine whether the learned strategy remains useful when the requests, core wording, or model differ from those used in training.

\subsection{Analysis of Optimization Quality}

Having examined where AMO improves visibility, we turn to the descriptions it produces.
We assess readability, source grounding, and recommendation salience to complete \textbf{RQ4}.

\begin{table}[!ht]
\centering
\caption{LLM-as-judge quality scores for rewritten descriptions. Avg. is the mean across dimensions.}
\label{tab:judge-quality-scores}
\begin{tabular}{lcccc}
\toprule
Method & Natural & Consist & Salience & Avg. \\
\midrule
Content Impr. & 4.82 & 4.21 & 4.83 & \textbf{4.62} \\
AMO & 4.71 & 4.01 & \textbf{4.99} & 4.57 \\
MetaOpt & 4.71 & 4.02 & 4.98 & 4.57 \\
AutoGEO & 4.59 & 4.18 & 4.82 & 4.53 \\
RecGEO & 4.54 & 4.03 & \textbf{4.99} & 4.52 \\
MetaOpt-Rec & 4.32 & 4.00 & \textbf{4.99} & 4.43 \\
\midrule
LLM Guid. & 4.09 & 4.42 & 4.50 & 4.34 \\
Fluency & 4.25 & \textbf{4.48} & 4.29 & 4.34 \\
\midrule
Quotes & 3.96 & 3.94 & 4.71 & 4.21 \\
Citations & 3.98 & 3.95 & 4.48 & 4.14 \\
Authoritative & 3.98 & 3.94 & 4.45 & 4.12 \\
Tech. Terms & 3.99 & 4.18 & 4.16 & 4.11 \\
Simple Lang. & 3.98 & 4.20 & 3.97 & 4.05 \\
Unique Words & 3.95 & 4.14 & 4.01 & 4.03 \\
Statistics & 3.87 & 2.98 & 3.90 & 3.58 \\
\bottomrule
\end{tabular}
\end{table}

\paragraph{Rubrics.}
The same LLM judge scores original--rewritten description pairs on a 1--5 scale, seeing only the two documents.
\textbf{Natural} measures fluency and readability.
\textbf{Consist} assesses item identity and source grounding, penalizing factual distortion and unsupported claims without separately scoring omissions.
\textbf{Salience} measures how clearly source-supported strengths inform a recommendation.
We use the main benchmark's three rewrites per method and report means in Table~\ref{tab:judge-quality-scores}, with the rubric in Appendix~\ref{app:rubric}.

\paragraph{AMO maintains readable, recommendation-focused descriptions.}
AMO's Natural score of 4.71 and Salience of 4.99 indicate readable, recommendation-focused descriptions (Table~\ref{tab:judge-quality-scores}).
Its Consist score is 4.01, below several baselines, and Content Improvement leads the average quality ranking.
The judge compares rewrites with their sources, while citation evaluation tests their use when recommending items for a particular request.
AMO, RecGEO, and MetaOpt-Rec tie on Salience yet yield different citation gains, illustrating why the two evaluations provide complementary evidence.

\paragraph{Generic GEO tactics can compromise description quality.}
Statistics, Quotes, and Citations were introduced as effective visibility strategies in general GEO~\cite{geo}. In this recommendation setting, however, Statistics receives the lowest Consist score (2.98), while Quotes, Citations, and Authoritative all score below 4 on both Natural and Consist. These results expose a risk when transferring generic credibility tactics to item descriptions: added authority cues need support from the item's actual evidence. Visibility optimization should therefore assess the quality of the description it promotes, alongside whether the assistant cites it.

\section{GEO toward Practice}
\label{sec:practical}

\begin{figure}[!ht]
\centering
\includegraphics[width=0.96\columnwidth]{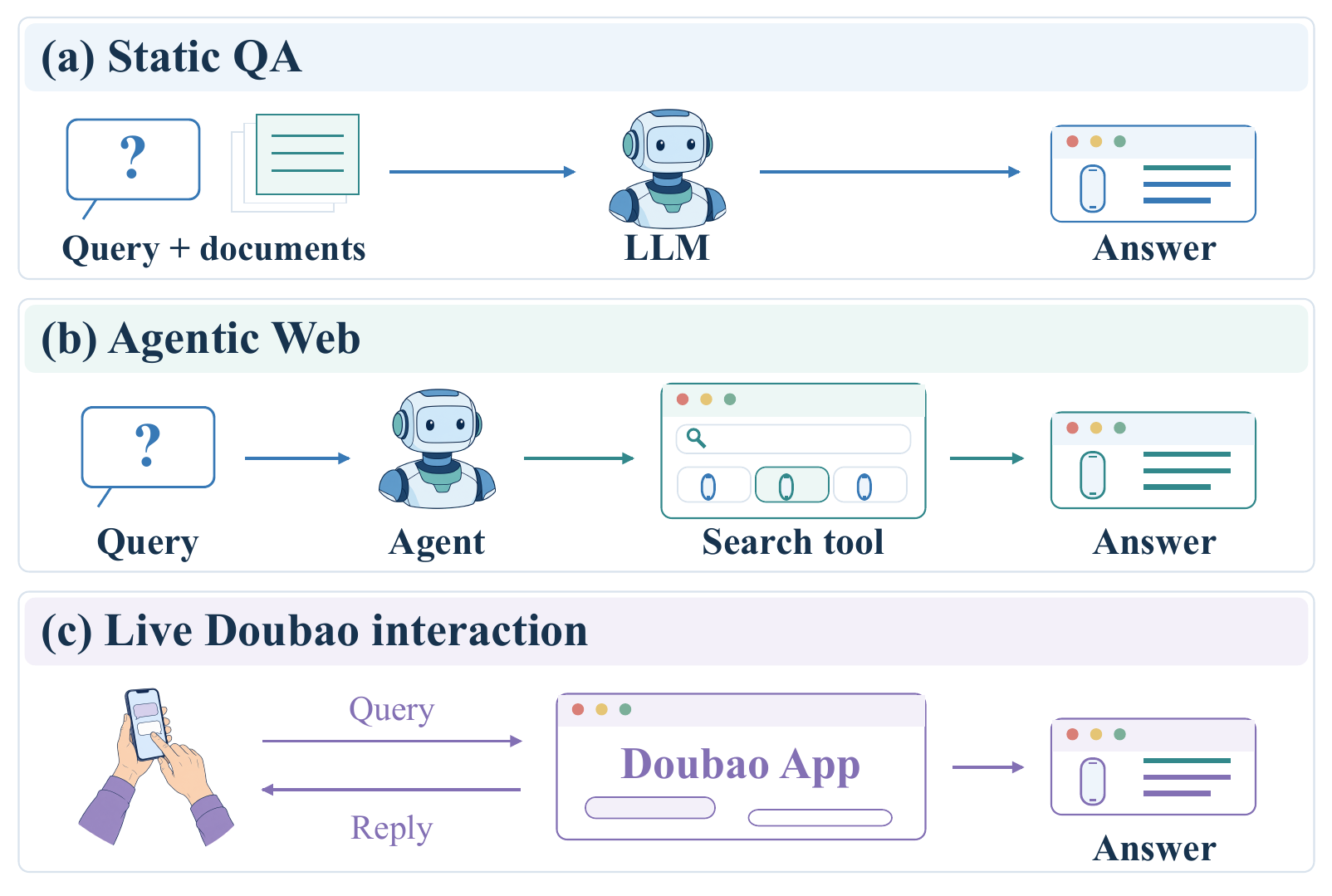}
\caption{\textbf{Generation-stage GEO across three assistant workflows.} Fixed candidate evidence is supplied directly (a), returned by the agent's search tool (b), or submitted through Doubao's official API with native search disabled (c).}
\Description{Three evaluation workflows. In the static setting, the model receives candidate descriptions directly. In the agentic setting, the assistant issues a search query, receives candidate evidence through a tool, and generates a recommendation. The third setting evaluates rewriting through the official Doubao assistant API.}
\label{fig:practical-contexts}
\end{figure}

The main experiments measure rewriting gains when candidate descriptions are supplied directly to an answer engine.
In practical recommendation settings, search tools mediate how evidence reaches the answer stage, while commercially deployed assistants operate under their own response rules~\cite{search_arena,kim2026sageo}.
For \textbf{RQ5}, we ask whether AMO's gains carry over to an agentic recommendation pipeline and whether it can adapt to a commercially deployed online assistant (Figure~\ref{fig:practical-contexts}).

\subsection{GEO in Agentic Recommendation}
\label{sec:agentic-context}
Agentic recommendation connects evidence gathering with the final choice.
The assistant translates a user's request into a search query, reads item descriptions returned by a search tool, and compares candidates to produce a cited recommendation.
We therefore reuse the saved benchmark rewrites to test whether their visibility gains carry over from direct context to this workflow without further optimization for the tool interface.

\paragraph{Setup.}
GPT-4o-mini receives the user request and issues one \texttt{web\_search} call with a query it generates.
The tool returns candidate descriptions and citation indices, after which the assistant produces its recommendation.
Candidate pools and input order are fixed across methods, retaining the generation-stage comparison within this agentic sequence.
We replace only the target description with each method's three saved benchmark rewrites and reuse them without retraining.
Appendix~\ref{app:practical} provides the protocol.

\begin{table}[ht]

\centering
\caption{Agentic recommendation without retraining. P/Q/All report mean rank gains over Base. Pass@5 is the top-five target citation rate.}
\label{tab:agentic-results}
\begin{tabular}{@{}lrrrr@{}}
\toprule
Method & P-Avg & Q-Avg & All & Pass@5 (\%) \\
\midrule
Base & --- & --- & --- & 33.85 \\
Content Impr. & 0.621 & 0.546 & 0.584 & 40.67 \\
RecGEO & 0.800 & 0.660 & 0.730 & 42.25 \\
AutoGEO & 0.599 & 0.482 & 0.540 & 40.72 \\
MetaOpt & 0.737 & 0.505 & 0.621 & 41.62 \\
\midrule
\textbf{AMO} & \textbf{0.929} & \textbf{0.734} & \textbf{0.832} & \textbf{43.48} \\
\bottomrule
\end{tabular}
\end{table}

\paragraph{AMO remains effective in agentic recommendation.}
AMO leads both intent families and Pass@5 when the assistant requests evidence through search (Table~\ref{tab:agentic-results}). RecGEO remains the strongest alternative, with the overall baseline ordering unchanged from the main benchmark. 
Their gains carry over from direct context to evidence requested during the recommendation task, connecting the shared-description setting to this agentic workflow without further prompt optimization.

\subsection{Optimization for the Doubao Assistant}
\label{sec:doubao-online}
We next extend the evaluation to a commercially deployed online assistant, bringing GEO closer to the services users consult for information and recommendations.
In these services, content visibility depends on how the assistant selects, cites, and presents item evidence under its own response rules.
We study Doubao through its official API, which exposes AI capabilities shared with the consumer Doubao App~\cite{volcengine_doubao_assistant}.
Its service-specific instructions and citation conventions provide a practical setting for examining whether AMO can adapt to the recommendation behavior of a deployed commercial assistant.

\paragraph{Setup.}
We access Doubao through Volcengine's official API~\cite{volcengine_doubao_assistant} (Figure~\ref{fig:practical-contexts}c), supplying candidate descriptions as textual search history with native search disabled.
All methods share requests and candidate pools across four domains and ten intent states.
We use DeepSeek-V4-Flash~\cite{DBLP:journals/corr/abs-2606-19348} as a strong backbone for rewriting and prompt updates, while Doubao provides citation feedback.
AMO and MetaOpt-Rec share the training and validation protocol.
Appendix~\ref{app:practical} gives the detailed setup.

\begin{table}[ht]

\centering
\caption{Doubao assistant evaluation with fixed evidence. P/Q/All: mean rank gains over Base; Pass@5: top-five target citation rate (\%).}
\label{tab:doubao-results}
\begin{tabular}{@{}lrrrr@{}}
\toprule
Method & P-Avg & Q-Avg & All & Pass@5 (\%) \\
\midrule
Base & --- & --- & --- & 43.52 \\
Content Impr. & 0.379 & 0.383 & 0.381 & 48.34 \\
RecGEO & 0.719 & 0.980 & 0.849 & 54.67 \\
AutoGEO & 0.746 & 1.065 & 0.906 & 56.68 \\
MetaOpt & 0.419 & 0.492 & 0.456 & 49.40 \\
MetaOpt-Rec & 0.710 & 0.873 & 0.791 & 54.02 \\
\midrule
\textbf{AMO} & \textbf{0.771} & \textbf{1.325} & \textbf{1.048} & \textbf{58.34} \\
\bottomrule
\end{tabular}
\end{table}

\paragraph{AMO achieves the largest mean gain on Doubao.}
The baseline ordering changes under Doubao's response rules, with AutoGEO moving ahead of RecGEO. AMO nevertheless achieves the largest mean gain on both intent families and the highest top-five citation rate among the evaluated methods (Table~\ref{tab:doubao-results}). Its advantage over MetaOpt-Rec under the shared training and validation protocol also supports adaptation to this assistant, complementing the reuse of benchmark rewrites in Section~\ref{sec:agentic-context}.

\begin{table}[ht]

\centering
\caption{Online Doubao: mean rank gains for Base-uncited ($|\mathcal U|=572$) and Base-cited ($|\mathcal C|=1{,}418$) cases. New citations report counts and percentages of $\mathcal U$.}
\label{tab:doubao-citation-strata}
\begin{tabular}{@{}lrrr@{}}
\toprule
Method & $\overline{\Delta R}_{\mathcal U}$ & $\overline{\Delta R}_{\mathcal C}$ & $N_{\mathrm{new}}$ (\%) \\
\midrule
Content Impr. & 1.470 & -0.059 & 149 (26.0) \\
RecGEO & 2.596 & 0.145 & 233 (40.7) \\
AutoGEO & 2.813 & 0.136 & 256 (44.8) \\
MetaOpt-Rec & 2.467 & 0.116 & 222 (38.8) \\
\textbf{AMO} & \textbf{2.846} & \textbf{0.322} & \textbf{264 (46.2)} \\
\bottomrule
\end{tabular}
\end{table}

\paragraph{AMO improves visibility from both starting conditions.}
We group requests by whether the saved Base answer cites the target (Table~\ref{tab:doubao-citation-strata}). For initially uncited targets, new citations show whether rewriting brings the item into the answer; for already cited targets, rank gains track changes from their initial visibility.
AMO improves mean rank in both groups.
Nearly half of initially uncited targets gain a citation, while already cited targets achieve the largest mean rank gain.
Content Improvement also gains new citations, but its rank gain is negative for targets already cited by Base.
New citations can therefore accompany declining mean visibility among items already present in the answers.

The matched investing-book example in Appendix~\ref{app:example} concerns a beginner seeking a single best choice.
The original description covers bear and bull markets, situation-specific guidance, and three-year real-world examples.
AMO brings these attributes together as reasons to choose the book, while MetaOpt lists the topics covered.
The two presentations give the assistant different ways to use the same item evidence in answering the final-choice request.
Doubao cites the AMO version first and leaves MetaOpt uncited.

\section{Related Work}

\paragraph{Black-hat GEO}
Black-hat GEO studies adversarial manipulation of search and ranking systems, where externally controlled content is crafted to steer model preferences or output rankings. Prompt injections and strategically optimized text can promote target items in conversational search and generative engines~\citep{ragdoll,black-hat-geo,kumar2024manipulating,tang2025stealthrank,DBLP:conf/emnlp/QianZJJW25,jin2026controlling}. 
These approaches exploit content as an instruction or ranking-control signal. 
Our white-hat setting instead requires natural, factual descriptions faithful to the original item, and studies whether recommendation evidence remains useful across intents.

\paragraph{White-hat GEO}
White-hat GEO seeks to improve a document's visibility while keeping its content natural and faithful to the source.
LLM-based white-hat SEO studies quality-preserving rewrites for search rankings~\cite{bardas2025white}, while related work considers content fidelity in ad retrieval and answer inclusion~\cite{ho2025rewrite}.
For generative engines, GEO~\cite{geo} and C-SEO Bench~\cite{c-seo} examine how rewriting affects source visibility in generated answers.
To guide rewriting, methods use inferred search goals to identify what the content should emphasize~\cite{raid_white}, or learn reusable rules reflecting engine preferences~\cite{autogeo}.
MetaOpt further uses visibility feedback to optimize the rewriting prompt~\cite{meta-optimization}.
By updating the instructions used for subsequent rewrites, prompt optimization learns a reusable strategy from engine responses.
Our work complements these approaches by evaluating shared rewrites across recommendation needs and decision stages.
AMO builds on feedback-driven prompt optimization while retaining recommendation instructions in a fixed core.

\section{Conclusion}
We study recommendation GEO under controlled changes in need specificity and decision progression. 
We find that strategies effective for broad requests can lose visibility when an item closely matches the stated need, and that unrestricted prompt optimization can erode the advantage of an effective recommendation seed.
To address this challenge, we propose AMO, which keeps recommendation instructions fixed while learning complementary rewriting strategies from citation feedback. 
AMO achieves the strongest aggregate visibility gains among the evaluated methods, with ablations showing that learned additions improve on the fixed core. 
These gains are accompanied by acceptable judged rewrite quality. 
The advantage also holds in the evaluated agentic workflow and, after separate optimization, on the deployed Doubao assistant. 
Together, these findings highlight the importance of assessing visibility across requests alongside how well a description presents the evidence an assistant needs to justify an item's fit.

\label{maintext-end}

\bibliographystyle{ACM-Reference-Format}
\bibliography{custom}

\appendix

\section{Data}
\label{sec:appendix-data}

\begin{table}[htbp]

\centering
\small
\caption{Dataset splits before intent expansion. Cand. denotes the average candidate pool size per test instance.}
\label{tab:data-splits}
\begin{tabular}{lrrrr}
\toprule
Dataset & Train & Val. & Test & Cand. \\
\midrule
Books & 174 & 25 & 50 & 9.02 \\
Retail & 200 & 25 & 50 & 10.00 \\
Videogames & 200 & 25 & 50 & 10.00 \\
Ragdoll & 0 & 0 & 50 & 8.00 \\
\midrule
\textbf{Total} & \textbf{574} & \textbf{75} & \textbf{200} & \textbf{9.26} \\
\bottomrule
\end{tabular}

\end{table}

Table~\ref{tab:data-splits} reports query-disjoint splits, with Ragdoll reserved for transfer evaluation.
Original-query training uses P2/Q3; P/Q-mixed training samples one state per example from the same training pool.

\section{Experimental Settings}
\label{app:param}
\label{app:practical}

Table~\ref{tab:appendix-settings} lists shared settings.
P2 and Q3 use identical wording but independently sampled answers.

\begin{table}[htbp]
\centering
\caption{Shared experimental settings for recommendation evaluation and prompt optimization.}
\label{tab:appendix-settings}
\resizebox{\linewidth}{!}{
\begin{tabular}{ll}
\toprule
\textbf{Setting} & \textbf{Value} \\
\midrule
Primary backbone & \texttt{gpt-4o-mini} \\
Rewrite / answer / query temperature & 0.7 / 0.7 / 0.3 \\
Top-$p$ / random seed & 1.0 / 42 \\
Epochs / batches per epoch / batch size & 3 / 12 / 50 \\
Rewrite variants per selected prompt & 3 \\
LLM judge / temperature & \texttt{gpt-4o-mini} / 0.0 \\
LLM judge context & before/after document only \\
\bottomrule
\end{tabular}
}
\end{table}

\paragraph{AMO optimization.}
\label{app:amo-optimization}
For training and validation, AMO averages MetaOpt's citation-conditioned rewards.
For $n$ candidates, with cited ranks $b$ before and $a$ after rewriting, rewards are $b-a$ for retained citations, $n-a$ for newly cited targets, and $-b$ for lost citations.
The mean $\widehat R(P_t;\mathcal B_t)$ excludes jointly uncited pairs and is zero if none remain; test results use all-instance $\Delta R$ (Section~\ref{sec:preliminary}).

Feedback $\mathcal F_t$ contains the mean, median, positive-reward rate, counts and histogram; $H_t$ records additions and feedback through round $t$.
The optimizer $M$ receives $A_t$, $\mathcal F_t$, $H_t$, and a read-only $\mathrm{summary}(C)$.
$\mathrm{ExtractAdditions}$ parses its proposal into $A_{t+1}$ for $P_{t+1}=[C;A_{t+1}]$.
We validate only updated prompts, retaining $A^\star$ whenever $s_{t+1}=\widehat R(P_{t+1};\mathcal D_{\mathrm{val}})>s^\star$, starting from $s^\star=-\infty$, and return $[C;A^\star]$.

\paragraph{Tool-mediated evidence}
The assistant uses the benchmark system instruction and sampling settings. It requests candidate evidence through a tool and then answers from the returned descriptions without further retrieval.

\paragraph{Online Doubao.}
We use \texttt{doubao-seed-\allowbreak 2-1-pro-\allowbreak 260628} with default assistant instructions.
Native search is disabled; all methods receive identical candidate evidence as textual search history.
Removing six unusable instances leaves 571 training, 73 validation, and 199 test instances.
AMO and MetaOpt-Rec train for one epoch with identical batches and validation rules; DeepSeek-V4-Flash generates rewrites and prompt updates.
Each method uses one rewrite and shared Base answers, scoring each state once and retaining uncited targets.
MetaOpt starts from Content Improvement.

\section{Rubrics}
\label{app:rubric}

The judge scores 600 pairs per method (200 instances, three rewrites) on a 1--5 scale, seeing only original and rewritten documents (Table~\ref{tab:judge-quality-scores}).

\paragraph{Natural.}
The rewrite is grammatical, coherent, and readable.

\paragraph{Consist.}
The rewrite preserves item identity and source grounding, without factual distortion or unsupported concrete claims.

\paragraph{Salience.}
Source-supported strengths make the item useful and compelling for recommendation.

\section{Full Results}
\label{app:full-results}

Tables~\ref{tab:appendix-full-p} and~\ref{tab:appendix-full-q} report all methods, including MetaOpt variants, with standard deviations and Pass@5.
AMO's mean gains are 0.98 on Books, 1.32 on Retail, 1.01 on Videogames, and 1.19 on unseen Ragdoll.
It exceeds RecGEO and MetaOpt-Prog in each domain; AutoGEO gains 1.05 on Videogames.
On Ragdoll, RecGEO and MetaOpt-Prog gain 0.90 and 0.97.
SDs are population standard deviations across three rewrite means; family averages first average each rewrite's five states.
Best and second-best results are bold and underlined.

\begin{table*}[t]
\centering
\caption{Full P-query results: mean $\pm$ population SD across three rewrites. Pass@5 in \%.}
\label{tab:appendix-full-p}
\resizebox{\textwidth}{!}{
\begin{tabular}{lcccccccccccc}
\toprule
\multirow{2}{*}{\textbf{Method}} & \multicolumn{2}{c}{\textbf{P1}} & \multicolumn{2}{c}{\textbf{P2}} & \multicolumn{2}{c}{\textbf{P3}} & \multicolumn{2}{c}{\textbf{P4}} & \multicolumn{2}{c}{\textbf{P5}} & \multicolumn{2}{c}{\textbf{Avg.}} \\
\cmidrule(lr){2-3}\cmidrule(lr){4-5}\cmidrule(lr){6-7}\cmidrule(lr){8-9}\cmidrule(lr){10-11}\cmidrule(lr){12-13}
 & \textbf{Pass@5} & \textbf{$\Delta$R} & \textbf{Pass@5} & \textbf{$\Delta$R} & \textbf{Pass@5} & \textbf{$\Delta$R} & \textbf{Pass@5} & \textbf{$\Delta$R} & \textbf{Pass@5} & \textbf{$\Delta$R} & \textbf{Pass@5} & \textbf{$\Delta$R} \\
\midrule
Authoritative & 27.7 $\pm$ 3.1& +1.36 $\pm$ 0.15& 24.3 $\pm$ 0.8& +0.54 $\pm$ 0.05& 39.5 $\pm$ 0.7& +0.21 $\pm$ 0.14& 58.5 $\pm$ 3.2& +0.40 $\pm$ 0.24& 86.7 $\pm$ 1.2& -0.12 $\pm$ 0.04& 47.3 $\pm$ 1.4& +0.48 $\pm$ 0.06 \\
Statistics & 24.0 $\pm$ 1.9& +1.19 $\pm$ 0.06& 23.5 $\pm$ 1.6& +0.48 $\pm$ 0.13& 38.3 $\pm$ 0.2& +0.21 $\pm$ 0.03& 56.0 $\pm$ 0.7& +0.24 $\pm$ 0.03& 86.5 $\pm$ 0.0& -0.18 $\pm$ 0.02& 45.7 $\pm$ 0.4& +0.39 $\pm$ 0.02 \\
Citations & 24.8 $\pm$ 1.3& +1.20 $\pm$ 0.14& 25.3 $\pm$ 0.9& +0.56 $\pm$ 0.10& 39.7 $\pm$ 3.1& +0.28 $\pm$ 0.19& 57.8 $\pm$ 1.7& +0.36 $\pm$ 0.11& 86.3 $\pm$ 1.6& -0.14 $\pm$ 0.13& 46.8 $\pm$ 0.6& +0.45 $\pm$ 0.07 \\
Fluency & 24.3 $\pm$ 1.2& +1.19 $\pm$ 0.02& 27.3 $\pm$ 0.8& +0.71 $\pm$ 0.03& 41.0 $\pm$ 1.1& +0.34 $\pm$ 0.06& 58.8 $\pm$ 2.4& +0.45 $\pm$ 0.18& 85.2 $\pm$ 0.5& -0.27 $\pm$ 0.04& 47.3 $\pm$ 0.3& +0.48 $\pm$ 0.03 \\
Unique Words & 22.5 $\pm$ 1.6& +0.99 $\pm$ 0.09& 21.5 $\pm$ 1.1& +0.35 $\pm$ 0.07& 36.8 $\pm$ 1.0& +0.03 $\pm$ 0.08& 52.3 $\pm$ 0.8& +0.01 $\pm$ 0.05& 85.2 $\pm$ 0.2& -0.26 $\pm$ 0.03& 43.7 $\pm$ 0.5& +0.22 $\pm$ 0.02 \\
Tech. Terms & 23.3 $\pm$ 3.1& +1.19 $\pm$ 0.20& 23.2 $\pm$ 0.8& +0.52 $\pm$ 0.07& 38.8 $\pm$ 2.5& +0.24 $\pm$ 0.21& 55.5 $\pm$ 1.1& +0.23 $\pm$ 0.07& 85.7 $\pm$ 0.6& -0.22 $\pm$ 0.02& 45.3 $\pm$ 0.3& +0.39 $\pm$ 0.04 \\
Simple Lang. & 24.5 $\pm$ 1.4& +1.18 $\pm$ 0.15& 25.7 $\pm$ 1.4& +0.53 $\pm$ 0.09& 38.8 $\pm$ 2.1& +0.17 $\pm$ 0.15& 54.3 $\pm$ 0.5& +0.09 $\pm$ 0.01& 84.3 $\pm$ 1.3& -0.37 $\pm$ 0.12& 45.5 $\pm$ 0.8& +0.32 $\pm$ 0.07 \\
Quotes & 29.8 $\pm$ 0.9& +1.48 $\pm$ 0.04& 26.8 $\pm$ 2.7& +0.69 $\pm$ 0.11& 41.2 $\pm$ 1.3& +0.40 $\pm$ 0.08& 57.5 $\pm$ 1.1& +0.38 $\pm$ 0.09& 85.8 $\pm$ 2.1& -0.18 $\pm$ 0.19& 48.2 $\pm$ 0.6& +0.55 $\pm$ 0.03 \\
LLM Guid. & 24.0 $\pm$ 1.8& +1.09 $\pm$ 0.08& 22.7 $\pm$ 1.5& +0.39 $\pm$ 0.09& 39.5 $\pm$ 1.2& +0.28 $\pm$ 0.05& 54.2 $\pm$ 0.8& +0.23 $\pm$ 0.14& 80.5 $\pm$ 0.7& -0.70 $\pm$ 0.06& 44.2 $\pm$ 0.7& +0.26 $\pm$ 0.02 \\
Content Impr. & 32.5 $\pm$ 2.0& +1.92 $\pm$ 0.14& 31.3 $\pm$ 1.4& +1.04 $\pm$ 0.09& 48.7 $\pm$ 1.2& +0.94 $\pm$ 0.14& 62.0 $\pm$ 0.7& +0.83 $\pm$ 0.14& 87.3 $\pm$ 2.0& -0.05 $\pm$ 0.17& 52.4 $\pm$ 0.8& +0.94 $\pm$ 0.07 \\
RecGEO & 36.2 $\pm$ 1.0& +2.12 $\pm$ 0.08& 35.0 $\pm$ 2.2& +1.31 $\pm$ 0.06& \cellcolor[rgb]{1.00,0.91,0.90}\textbf{52.5 $\pm$ 0.7}& \cellcolor[rgb]{1.00,0.91,0.90}\textbf{+1.26 $\pm$ 0.07}& \cellcolor[gray]{0.94}\underline{64.2 $\pm$ 0.8}& \cellcolor[gray]{0.94}\underline{+1.00 $\pm$ 0.07}& \cellcolor[gray]{0.94}\underline{87.7 $\pm$ 0.6}& \cellcolor[gray]{0.94}\underline{+0.01 $\pm$ 0.05}& \cellcolor[gray]{0.94}\underline{55.1 $\pm$ 0.4}& \cellcolor[gray]{0.94}\underline{+1.14 $\pm$ 0.01} \\
\midrule
AutoGEO & 31.5 $\pm$ 2.9& +1.88 $\pm$ 0.13& 35.0 $\pm$ 1.9& +1.34 $\pm$ 0.11& \cellcolor[gray]{0.94}\underline{50.5 $\pm$ 1.1}& \cellcolor[gray]{0.94}\underline{+1.12 $\pm$ 0.03}& 62.3 $\pm$ 4.2& +0.76 $\pm$ 0.26& 84.7 $\pm$ 1.4& -0.29 $\pm$ 0.10& 52.8 $\pm$ 1.7& +0.96 $\pm$ 0.10 \\
MetaOpt & 35.8 $\pm$ 0.2& +2.08 $\pm$ 0.03& 33.0 $\pm$ 3.3& +1.17 $\pm$ 0.17& 50.2 $\pm$ 1.7& +1.05 $\pm$ 0.12& 61.7 $\pm$ 0.6& +0.69 $\pm$ 0.04& 83.2 $\pm$ 1.6& -0.42 $\pm$ 0.13& 52.8 $\pm$ 1.2& +0.91 $\pm$ 0.08 \\
MetaOpt-Rec & \cellcolor[gray]{0.94}\underline{38.2 $\pm$ 1.9}& \cellcolor[gray]{0.94}\underline{+2.26 $\pm$ 0.15}& 34.5 $\pm$ 2.2& +1.26 $\pm$ 0.16& 45.0 $\pm$ 1.1& +0.67 $\pm$ 0.10& 57.5 $\pm$ 0.7& +0.43 $\pm$ 0.05& 81.0 $\pm$ 0.0& -0.66 $\pm$ 0.03& 51.2 $\pm$ 0.8& +0.79 $\pm$ 0.04 \\
MetaOpt-Prog & 36.2 $\pm$ 1.0& +2.12 $\pm$ 0.06& \cellcolor[gray]{0.94}\underline{35.8 $\pm$ 1.3}& \cellcolor[gray]{0.94}\underline{+1.38 $\pm$ 0.08}& 49.2 $\pm$ 0.8& +1.05 $\pm$ 0.02& 62.8 $\pm$ 3.0& +0.79 $\pm$ 0.27& 82.7 $\pm$ 0.2& -0.42 $\pm$ 0.04& 53.3 $\pm$ 0.8& +0.98 $\pm$ 0.08 \\
MetaOpt-Rec-Prog & \cellcolor[gray]{0.94}\underline{38.2 $\pm$ 2.6}& +2.21 $\pm$ 0.21& 35.2 $\pm$ 0.2& \cellcolor[gray]{0.94}\underline{+1.38 $\pm$ 0.03}& 45.5 $\pm$ 0.4& +0.76 $\pm$ 0.09& 57.8 $\pm$ 1.3& +0.47 $\pm$ 0.12& 81.7 $\pm$ 1.3& -0.58 $\pm$ 0.13& 51.7 $\pm$ 0.2& +0.85 $\pm$ 0.05 \\
\midrule
AMO & \cellcolor[rgb]{1.00,0.91,0.90}\textbf{40.8 $\pm$ 2.9}& \cellcolor[rgb]{1.00,0.91,0.90}\textbf{+2.44 $\pm$ 0.14}& \cellcolor[rgb]{1.00,0.91,0.90}\textbf{36.7 $\pm$ 2.1}& \cellcolor[rgb]{1.00,0.91,0.90}\textbf{+1.46 $\pm$ 0.06}& 47.8 $\pm$ 3.0& +1.03 $\pm$ 0.15& \cellcolor[rgb]{1.00,0.91,0.90}\textbf{66.0 $\pm$ 0.4}& \cellcolor[rgb]{1.00,0.91,0.90}\textbf{+1.13 $\pm$ 0.07}& \cellcolor[rgb]{1.00,0.91,0.90}\textbf{88.3 $\pm$ 2.1}& \cellcolor[rgb]{1.00,0.91,0.90}\textbf{+0.06 $\pm$ 0.21}& \cellcolor[rgb]{1.00,0.91,0.90}\textbf{55.9 $\pm$ 1.2}& \cellcolor[rgb]{1.00,0.91,0.90}\textbf{+1.22 $\pm$ 0.08} \\
\bottomrule
\end{tabular}}

\centering
\caption{Full Q-query results: mean $\pm$ population SD across three rewrites. Pass@5 in \%.}
\label{tab:appendix-full-q}
\resizebox{\textwidth}{!}{
\begin{tabular}{lcccccccccccc}
\toprule
\multirow{2}{*}{\textbf{Method}} & \multicolumn{2}{c}{\textbf{Q1}} & \multicolumn{2}{c}{\textbf{Q2}} & \multicolumn{2}{c}{\textbf{Q3}} & \multicolumn{2}{c}{\textbf{Q4}} & \multicolumn{2}{c}{\textbf{Q5}} & \multicolumn{2}{c}{\textbf{Avg.}} \\
\cmidrule(lr){2-3}\cmidrule(lr){4-5}\cmidrule(lr){6-7}\cmidrule(lr){8-9}\cmidrule(lr){10-11}\cmidrule(lr){12-13}
 & \textbf{Pass@5} & \textbf{$\Delta$R} & \textbf{Pass@5} & \textbf{$\Delta$R} & \textbf{Pass@5} & \textbf{$\Delta$R} & \textbf{Pass@5} & \textbf{$\Delta$R} & \textbf{Pass@5} & \textbf{$\Delta$R} & \textbf{Pass@5} & \textbf{$\Delta$R} \\
\midrule
Authoritative & 27.5 $\pm$ 0.4& +0.00 $\pm$ 0.06& 26.7 $\pm$ 0.8& +0.17 $\pm$ 0.01& 24.0 $\pm$ 1.5& +0.14 $\pm$ 0.09& 26.2 $\pm$ 1.7& +0.15 $\pm$ 0.11& 18.0 $\pm$ 1.5& +0.35 $\pm$ 0.10& 24.5 $\pm$ 1.0& +0.16 $\pm$ 0.06 \\
Statistics & 26.7 $\pm$ 1.0& -0.04 $\pm$ 0.12& 26.7 $\pm$ 0.6& +0.19 $\pm$ 0.09& 19.3 $\pm$ 0.2& -0.14 $\pm$ 0.07& 25.3 $\pm$ 1.0& +0.07 $\pm$ 0.04& 17.7 $\pm$ 0.6& +0.36 $\pm$ 0.08& 23.1 $\pm$ 0.4& +0.09 $\pm$ 0.05 \\
Citations & 28.2 $\pm$ 0.9& +0.05 $\pm$ 0.08& 26.5 $\pm$ 1.1& +0.19 $\pm$ 0.11& 22.7 $\pm$ 2.4& +0.11 $\pm$ 0.19& 26.3 $\pm$ 2.1& +0.21 $\pm$ 0.08& 16.5 $\pm$ 0.4& +0.30 $\pm$ 0.02& 24.0 $\pm$ 0.2& +0.17 $\pm$ 0.05 \\
Fluency & 27.7 $\pm$ 1.2& +0.14 $\pm$ 0.05& 28.8 $\pm$ 1.3& +0.25 $\pm$ 0.04& 25.5 $\pm$ 2.5& +0.28 $\pm$ 0.14& 27.3 $\pm$ 1.3& +0.24 $\pm$ 0.23& 15.8 $\pm$ 1.0& +0.27 $\pm$ 0.10& 25.0 $\pm$ 0.3& +0.23 $\pm$ 0.08 \\
Unique Words & 24.8 $\pm$ 0.9& -0.15 $\pm$ 0.03& 26.0 $\pm$ 0.7& +0.12 $\pm$ 0.03& 22.0 $\pm$ 1.5& -0.05 $\pm$ 0.10& 25.8 $\pm$ 0.8& +0.07 $\pm$ 0.02& 13.3 $\pm$ 1.0& +0.06 $\pm$ 0.08& 22.4 $\pm$ 0.4& +0.01 $\pm$ 0.01 \\
Tech. Terms & 27.2 $\pm$ 0.8& +0.03 $\pm$ 0.02& 27.2 $\pm$ 0.6& +0.21 $\pm$ 0.04& 22.0 $\pm$ 0.7& +0.09 $\pm$ 0.18& 26.2 $\pm$ 1.3& +0.07 $\pm$ 0.08& 14.8 $\pm$ 1.2& +0.21 $\pm$ 0.08& 23.5 $\pm$ 0.4& +0.12 $\pm$ 0.03 \\
Simple Lang. & 27.2 $\pm$ 0.8& +0.07 $\pm$ 0.03& 28.0 $\pm$ 1.8& +0.22 $\pm$ 0.16& 22.5 $\pm$ 2.9& +0.07 $\pm$ 0.11& 26.8 $\pm$ 2.1& +0.15 $\pm$ 0.04& 14.3 $\pm$ 1.2& +0.13 $\pm$ 0.11& 23.8 $\pm$ 0.6& +0.13 $\pm$ 0.05 \\
Quotes & 29.7 $\pm$ 0.8& +0.15 $\pm$ 0.05& 28.5 $\pm$ 0.8& +0.30 $\pm$ 0.03& 22.7 $\pm$ 1.9& +0.10 $\pm$ 0.07& 25.8 $\pm$ 0.5& +0.17 $\pm$ 0.01& 17.7 $\pm$ 1.7& +0.43 $\pm$ 0.18& 24.9 $\pm$ 0.6& +0.23 $\pm$ 0.04 \\
LLM Guid. & 24.5 $\pm$ 1.2& -0.52 $\pm$ 0.09& 23.5 $\pm$ 0.8& -0.28 $\pm$ 0.06& 22.5 $\pm$ 2.2& +0.01 $\pm$ 0.17& 23.0 $\pm$ 0.7& -0.16 $\pm$ 0.08& 16.3 $\pm$ 2.9& +0.29 $\pm$ 0.16& 22.0 $\pm$ 1.0& -0.13 $\pm$ 0.06 \\
Content Impr. & 31.5 $\pm$ 1.1& \cellcolor[gray]{0.94}\underline{+0.42 $\pm$ 0.06}& 32.3 $\pm$ 0.2& +0.57 $\pm$ 0.03& 31.3 $\pm$ 1.0& +0.72 $\pm$ 0.08& 32.7 $\pm$ 0.6& +0.66 $\pm$ 0.01& 23.5 $\pm$ 1.1& +0.86 $\pm$ 0.07& 30.3 $\pm$ 0.5& +0.65 $\pm$ 0.02 \\
RecGEO & 31.8 $\pm$ 1.5& +0.35 $\pm$ 0.03& \cellcolor[gray]{0.94}\underline{33.5 $\pm$ 1.5}& \cellcolor[gray]{0.94}\underline{+0.75 $\pm$ 0.07}& \cellcolor[gray]{0.94}\underline{33.7 $\pm$ 1.5}& +0.83 $\pm$ 0.11& \cellcolor[gray]{0.94}\underline{33.3 $\pm$ 0.5}& \cellcolor[gray]{0.94}\underline{+0.80 $\pm$ 0.04}& 27.2 $\pm$ 1.0& +1.11 $\pm$ 0.07& \cellcolor[gray]{0.94}\underline{31.9 $\pm$ 0.5}& \cellcolor[gray]{0.94}\underline{+0.77 $\pm$ 0.04} \\
\midrule
AutoGEO & 33.5 $\pm$ 1.2& +0.36 $\pm$ 0.02& 30.8 $\pm$ 2.1& +0.43 $\pm$ 0.12& 32.8 $\pm$ 0.2& +0.81 $\pm$ 0.04& 30.2 $\pm$ 0.9& +0.52 $\pm$ 0.08& 22.7 $\pm$ 0.6& +0.79 $\pm$ 0.05& 30.0 $\pm$ 0.5& +0.58 $\pm$ 0.05 \\
MetaOpt & 30.2 $\pm$ 1.4& +0.14 $\pm$ 0.05& 31.0 $\pm$ 1.2& +0.54 $\pm$ 0.11& 32.8 $\pm$ 1.9& +0.83 $\pm$ 0.09& 33.0 $\pm$ 1.4& +0.73 $\pm$ 0.05& \cellcolor[gray]{0.94}\underline{27.7 $\pm$ 0.6}& \cellcolor[gray]{0.94}\underline{+1.16 $\pm$ 0.05}& 30.9 $\pm$ 0.5& +0.68 $\pm$ 0.03 \\
MetaOpt-Rec & 29.8 $\pm$ 1.3& +0.05 $\pm$ 0.07& 30.0 $\pm$ 0.7& +0.28 $\pm$ 0.09& 31.7 $\pm$ 1.6& +0.81 $\pm$ 0.11& 33.0 $\pm$ 0.8& +0.58 $\pm$ 0.11& 26.2 $\pm$ 1.4& +1.05 $\pm$ 0.06& 30.1 $\pm$ 0.6& +0.56 $\pm$ 0.03 \\
MetaOpt-Prog & \cellcolor[gray]{0.94}\underline{33.7 $\pm$ 1.3}& +0.39 $\pm$ 0.10& 30.2 $\pm$ 1.3& +0.49 $\pm$ 0.05& 33.3 $\pm$ 2.1& \cellcolor[gray]{0.94}\underline{+0.85 $\pm$ 0.11}& 33.2 $\pm$ 2.2& +0.68 $\pm$ 0.11& 25.0 $\pm$ 1.6& +0.98 $\pm$ 0.15& 31.1 $\pm$ 1.0& +0.68 $\pm$ 0.05 \\
MetaOpt-Rec-Prog & 30.0 $\pm$ 2.0& +0.12 $\pm$ 0.04& 29.3 $\pm$ 1.2& +0.29 $\pm$ 0.14& 31.2 $\pm$ 3.2& +0.74 $\pm$ 0.14& 31.7 $\pm$ 2.5& +0.56 $\pm$ 0.11& 27.3 $\pm$ 1.0& +1.15 $\pm$ 0.07& 29.9 $\pm$ 0.9& +0.57 $\pm$ 0.03 \\
\midrule
AMO & \cellcolor[rgb]{1.00,0.91,0.90}\textbf{34.3 $\pm$ 1.7}& \cellcolor[rgb]{1.00,0.91,0.90}\textbf{+0.49 $\pm$ 0.09}& \cellcolor[rgb]{1.00,0.91,0.90}\textbf{34.3 $\pm$ 0.2}& \cellcolor[rgb]{1.00,0.91,0.90}\textbf{+0.78 $\pm$ 0.01}& \cellcolor[rgb]{1.00,0.91,0.90}\textbf{38.0 $\pm$ 1.8}& \cellcolor[rgb]{1.00,0.91,0.90}\textbf{+1.20 $\pm$ 0.08}& \cellcolor[rgb]{1.00,0.91,0.90}\textbf{37.8 $\pm$ 2.1}& \cellcolor[rgb]{1.00,0.91,0.90}\textbf{+1.06 $\pm$ 0.17}& \cellcolor[rgb]{1.00,0.91,0.90}\textbf{33.2 $\pm$ 1.7}& \cellcolor[rgb]{1.00,0.91,0.90}\textbf{+1.59 $\pm$ 0.09}& \cellcolor[rgb]{1.00,0.91,0.90}\textbf{35.5 $\pm$ 1.1}& \cellcolor[rgb]{1.00,0.91,0.90}\textbf{+1.02 $\pm$ 0.06} \\
\bottomrule
\end{tabular}}

\centering
\caption{P-query component ablations: mean $\pm$ population standard deviation over three rewrites. Pass@5 is a percentage.}
\label{tab:appendix-ablation-p}
\resizebox{\textwidth}{!}{\begin{tabular}{lrrrrrrrrrrrr}
\toprule
Method & \multicolumn{2}{c}{P1} & \multicolumn{2}{c}{P2} & \multicolumn{2}{c}{P3} & \multicolumn{2}{c}{P4} & \multicolumn{2}{c}{P5} & \multicolumn{2}{c}{P-Avg} \\
 & Pass@5 & $\Delta R$ & Pass@5 & $\Delta R$ & Pass@5 & $\Delta R$ & Pass@5 & $\Delta R$ & Pass@5 & $\Delta R$ & Pass@5 & $\Delta R$ \\
\midrule
Core only & 35.3 $\pm$ 1.7 & +1.98 $\pm$ 0.13 & 31.7 $\pm$ 2.2 & +1.11 $\pm$ 0.16 & 53.0 $\pm$ 0.8 & +1.29 $\pm$ 0.09 & 66.8 $\pm$ 0.2 & +1.17 $\pm$ 0.05 & 90.0 $\pm$ 2.0 & +0.23 $\pm$ 0.15 & 55.4 $\pm$ 0.9 & +1.15 $\pm$ 0.07 \\
Full-prompt opt. & 35.3 $\pm$ 2.1 & +1.92 $\pm$ 0.13 & 33.2 $\pm$ 3.1 & +1.17 $\pm$ 0.16 & 45.5 $\pm$ 3.1 & +0.80 $\pm$ 0.24 & 61.3 $\pm$ 1.7 & +0.71 $\pm$ 0.09 & 83.2 $\pm$ 0.6 & -0.41 $\pm$ 0.07 & 51.7 $\pm$ 1.2 & +0.84 $\pm$ 0.07 \\
AMO & 40.8 $\pm$ 2.9 & +2.44 $\pm$ 0.14 & 36.7 $\pm$ 2.1 & +1.46 $\pm$ 0.06 & 47.8 $\pm$ 3.0 & +1.03 $\pm$ 0.15 & 66.0 $\pm$ 0.4 & +1.13 $\pm$ 0.07 & 88.3 $\pm$ 2.1 & +0.06 $\pm$ 0.21 & 55.9 $\pm$ 1.2 & +1.22 $\pm$ 0.08 \\
\bottomrule
\end{tabular}}

\centering
\caption{Q-query component ablations: mean $\pm$ population standard deviation over three rewrites. Pass@5 is a percentage.}
\label{tab:appendix-ablation-q}
\resizebox{\textwidth}{!}{\begin{tabular}{lrrrrrrrrrrrr}
\toprule
Method & \multicolumn{2}{c}{Q1} & \multicolumn{2}{c}{Q2} & \multicolumn{2}{c}{Q3} & \multicolumn{2}{c}{Q4} & \multicolumn{2}{c}{Q5} & \multicolumn{2}{c}{Q-Avg} \\
 & Pass@5 & $\Delta R$ & Pass@5 & $\Delta R$ & Pass@5 & $\Delta R$ & Pass@5 & $\Delta R$ & Pass@5 & $\Delta R$ & Pass@5 & $\Delta R$ \\
\midrule
Core only & 32.8 $\pm$ 0.5 & +0.44 $\pm$ 0.08 & 32.2 $\pm$ 0.6 & +0.66 $\pm$ 0.05 & 34.3 $\pm$ 0.6 & +0.90 $\pm$ 0.07 & 33.5 $\pm$ 1.6 & +0.91 $\pm$ 0.11 & 26.7 $\pm$ 0.5 & +1.09 $\pm$ 0.02 & 31.9 $\pm$ 0.3 & +0.80 $\pm$ 0.03 \\
Full-prompt opt. & 30.0 $\pm$ 2.9 & +0.18 $\pm$ 0.09 & 30.2 $\pm$ 0.8 & +0.46 $\pm$ 0.06 & 30.7 $\pm$ 1.7 & +0.75 $\pm$ 0.18 & 32.3 $\pm$ 0.8 & +0.62 $\pm$ 0.04 & 28.2 $\pm$ 0.6 & +1.22 $\pm$ 0.05 & 30.3 $\pm$ 1.1 & +0.64 $\pm$ 0.06 \\
AMO & 34.3 $\pm$ 1.7 & +0.49 $\pm$ 0.09 & 34.3 $\pm$ 0.2 & +0.78 $\pm$ 0.01 & 38.0 $\pm$ 1.8 & +1.20 $\pm$ 0.08 & 37.8 $\pm$ 2.1 & +1.06 $\pm$ 0.17 & 33.2 $\pm$ 1.7 & +1.59 $\pm$ 0.09 & 35.5 $\pm$ 1.1 & +1.02 $\pm$ 0.06 \\
\bottomrule
\end{tabular}}
\end{table*}

\section{AMO Ablation Results}
\label{app:ablation}
Tables~\ref{tab:appendix-ablation-p} and~\ref{tab:appendix-ablation-q} share instances, states, backbone, and three rewrites per method.
Both optimizers train and validate on topic-level requests for three epochs (36 batches).
AMO selects among updated $[C;A]$ prompts using citation-conditioned reward (Appendix~\ref{app:amo-optimization}); full-prompt optimization uses all-instance $\Delta R$ and validates before updating, including the initial core.
Core only is evaluated separately.
All test results use all-instance $\Delta R$ (Section~\ref{sec:preliminary}).

\section{Qualitative Example}
\label{app:example}
For a beginner choosing an investing book, AMO links coverage to reasons for choosing it, whereas MetaOpt lists topics (Table~\ref{tab:amo-example}).

\begin{table}[htbp]

\centering
\caption{Online Doubao: AMO first, MetaOpt uncited.}
\label{tab:amo-example}
\footnotesize
\begin{tabular}{@{}p{0.17\columnwidth}p{\dimexpr0.83\columnwidth-2\tabcolsep\relax}@{}}
\toprule
\textbf{Field} & \textbf{Example} \\
\midrule
Query & recommend me the very best book about Investing for beginners \\

Original & Stock Investing For Dummies, 3rd Edition includes information on stock investing in both bear and bull markets; unique investment segments; stock investing for different types of situations; and examples straight from the real world of stock investing as they have occurred in the past three years. \\

MetaOpt & Stock Investing For Dummies, 3rd Edition includes information on stock investing in both bear markets and bull markets. Stock Investing For Dummies, 3rd Edition includes information on unique investment segments. \\

AMO & \textbf{Why Stock Investing For Dummies, 3rd Edition is recommended:} Stock Investing For Dummies, 3rd Edition is recommended because it combines dual market-regime coverage (bear and bull), unique investment segments, situation-specific guidance, and three-year real-world examples in a single volume \ldots \\

{\raggedright{}Service answer\par} & For new investors seeking a comprehensive, all-market guide, \textbf{Stock Investing For Dummies, 3rd Edition} stands out, \ldots{} covers strategies for both bull and bear markets, \ldots{} [5]. \\

Rank & Base: uncited; MetaOpt: uncited; \textbf{AMO: first cited} \\
\bottomrule
\end{tabular}

\end{table}

\section{Top-K Metrics and Wording Robustness}
\label{app:robustness}
AMO leads on all rescored MRR and Pass@$K$ metrics (Table~\ref{tab:topk}).

For a cited target at rank $r$, $\mathrm{RR}@K=\mathbf{1}[r\leq K]/r$; an uncited target receives zero. $\Delta\mathrm{MRR}@K$ averages the difference from Base. Pass@$K$ is the fraction of answers citing the target within the first $K$ unique citations; $\Delta$Pass@$K$ is its change in percentage points. Table~\ref{tab:topk} uses the full benchmark's three rewrites.

\begin{table}[htbp]

\centering
\caption{Top-K gains from Base. Pass: percentage points.}
\label{tab:topk}
\small
\begin{tabular}{lrrrrr}
\toprule
& \multicolumn{2}{c}{$\Delta$MRR} & \multicolumn{3}{c}{$\Delta$Pass}\\
Method & @5 & @10 & @1 & @3 & @5\\
\midrule
AMO & \textbf{.0720} & \textbf{.0752} & \textbf{4.03} & \textbf{9.13} & \textbf{13.98}\\
RecGEO & .0599 & .0635 & 3.58 & 6.52 & 11.75\\
MetaOpt-Prog & .0522 & .0546 & 2.72 & 6.73 & 10.45\\
AutoGEO & .0522 & .0537 & 2.90 & 6.50 & 9.65\\
Content Impr. & .0516 & .0542 & 3.12 & 6.20 & 9.57\\
\bottomrule
\end{tabular}

\end{table}

\begin{table}[!htbp]

\centering
\caption{Anchor-paraphrase robustness.
P-Avg and Q-Avg report mean $\Delta$Pass@5 over P1--P5 and Q1--Q5, respectively (percentage points over Base).}
\label{tab:anchor-paraphrase}
\begin{tabular}{lrr}
\toprule
Method & P-Avg & Q-Avg \\
\midrule
AMO (original anchor) & 12.25 & 13.25 \\
AMO (paraphrased anchor) & \textbf{15.50} & \textbf{13.50} \\
\midrule
RecGEO & 14.25 & 9.75 \\
MetaOpt-Prog & 13.75 & 8.25 \\
AutoGEO & 8.25 & 7.25 \\
\bottomrule
\end{tabular}

\end{table}

\begin{table}[!htbp]

\centering
\caption{Mean rank gain $\Delta R$ over Base across all intent states and three rewrites. Gemini and Qwen denote Gemini 2.5 Flash-Lite and Qwen3-30B-A3B, respectively.}
\label{tab:backbone-sensitivity}
\begin{tabular}{lrrr}
\toprule
Method & GPT-4o-mini & Gemini & Qwen \\
\midrule
Content Impr. & 0.79 & 0.41 & 0.48 \\
RecGEO & 0.95 & 0.54 & 0.44 \\
AutoGEO & 0.77 & 0.41 & 1.10 \\
MetaOpt-Prog & 0.83 & 0.50 & 0.65 \\
\midrule
AMO & \textbf{1.12} & \textbf{0.62} & \textbf{1.54} \\
\bottomrule
\end{tabular}

\end{table}

\paragraph{Natural query phrasing.}
We test template sensitivity by paraphrasing P3--P5 for all 200 test instances, preserving requirements, targets, and candidates.
Table~\ref{tab:query-paraphrase} reports Pass@5 gains over Base from one rewrite per method.

\paragraph{Alternative anchor wording.}
GPT-4o-mini paraphrases the core as an informed recommendation with concrete differentiators. Keeping the learned additions, we compare both wordings with the same baselines on 20 instances per domain, using one rewrite per method (Table~\ref{tab:anchor-paraphrase}).

\paragraph{Backbone transfer.}
For Table~\ref{tab:backbone-sensitivity}, each named model generates both rewrites and answers using the selected instructions and the same queries, targets, and candidate pools.

\section{Paired Evidence for Aggregate Gains}
\label{sec:paired-evidence}\label{app:uncertainty}
Table~\ref{tab:paired-intervals} reports the paired comparisons.

Within each domain, we bootstrap original test instances 20,000 times, retaining their states and rewrites, to obtain 95\% percentile intervals.
Paired comparisons cancel the shared Base rank. Intervals summarize answers from selected prompts.

\begin{table}[t]

\centering
\caption{Paired rank comparisons and 95\% confidence intervals. Main gains average all states unless marked P5; Base compares unrounded saved P2/Q3 ranks.}
\label{tab:paired-intervals}
\begin{tabular}{@{}llr@{}}
\toprule
Setting & Contrast & Difference [95\% CI] \\
\midrule
Main & AMO--RecGEO & $+0.172\;[+0.072,+0.271]$ \\
Main & AMO--MetaOpt-Prog & $+0.293\;[+0.175,+0.412]$ \\
Main & AMO--Core only & $+0.147\;[+0.056,+0.236]$ \\
Main & AMO--Full-prompt opt. & $+0.384\;[+0.268,+0.502]$ \\
Main P5 & AMO--Base & $+0.063\;[-0.305,+0.435]$ \\
Base & P2--Q3 & $+0.290\;[-0.040,+0.635]$ \\
Doubao & AMO--RecGEO & $+0.198\;[+0.067,+0.327]$ \\
Doubao & AMO--AutoGEO & $+0.142\;[-0.027,+0.298]$ \\
\bottomrule
\end{tabular}
\end{table}

\section{Prompt Templates}
\label{app:prompt}
\paragraph{Target-grounded query construction.}
P1 names the domain; P2/Q3 use ``[article] [domain] [about/for] [topic]''. Given the topic and candidate previews, the generator expresses one, two, and more target-supported attributes as requirements for P3--P5, keeping the target eligible without copying phrases. Queries are fixed before rewriting and shared by methods. Q1/Q2 request comprehensive/brief surveys; Q4 requests several items; Q5 asks for the single best choice.

\paragraph{Prompt-drift evidence.}
\label{app:drift-markers}
We inspect 36 saved prompts for literal markers, normalizing case and whitespace before counting. Retention and recommender markers match the seed phrases in Section~\ref{sec:core-repair}; keyword and call-to-action markers allow singular/plural and space/hyphen variants.

\paragraph{Frozen recommendation core.}
The source is a separate input.
\begingroup
\begin{lstlisting}[style=promptstyle]
1. Write as a knowledgeable domain expert making a confident, persuasive recommendation
2. Emphasize what makes this item uniquely valuable and superior to alternatives
3. Highlight strengths, advantages, and standout qualities that compel recommendation
4. Order by distinctiveness — lead with the most unique, differentiating features
5. Add authoritative content that an AI system would be more likely to cite and surface
6. Preserve ALL factual information — every name, number, spec, endorsement must appear
7. Every sentence must carry new information — no generic filler
Do not explain the changes, just give me the revised source.
\end{lstlisting}
\endgroup

\paragraph{Selected additions.}
The selected $A^\star$ appends the following instructions to the core.
\begingroup
\begin{lstlisting}[style=promptstyle]
Create unified sections for related features, highlighting collective advantages with clear headings.
Add actionable insights from recent trends or user feedback to enhance content relevance.
Use bullet points for quick reference of key differentiators to improve readability.
Include comparative analyses against competitors to reinforce claims of superiority.
Employ user-centric language that directly addresses potential queries or concerns.
Start each paragraph with a strong topic sentence to improve structural clarity.
Incorporate real-world examples or case studies to illustrate the benefits and add credibility.
\end{lstlisting}
\endgroup

\end{document}